\documentclass[11pt,twoside]{article}

\usepackage{asp2006}
\usepackage{epsf}
\usepackage{psfig}
\usepackage{graphicx}
\usepackage{lscape}

\begin{document}
\title{A Search for EHB Pulsators in the Globular Cluster NGC~6752}   %%% Fill in title
\author{M. Catelan, G. E. Prieto\altaffilmark{1}, 
        M. Zoccali, C. Weidner}   
\affil{P. Universidad Cat\'{o}lica de Chile,
       Dept. de Astronom\'{i}a y Astrof\'{i}sica, Av. Vicu\~{n}a Mackenna
       4860, 782-0436 Macul, Santiago, Chile}

\author{P. B. Stetson}   
\affil{Dominion Astrophysical Observatory, Herzberg Institute of Astrophysics, %Natl. Research Council, 
       5071 West Saanich Road, Victoria, BC V9E 2E7, Canada}
       
\author{C. Moni Bidin, M. Altmann\altaffilmark{2}}   
\affil{Departamento de Astronom\'{i}a, Universidad de Chile,
       Casilla 36-D, Santiago, Chile}

\author{H. A. Smith}   
\affil{Department of Physics and Astronomy, Michigan State University, 
       East Lansing, MI 48824, USA} 
       
\author{B. J. Pritzl}   
\affil{Department of Physics and Astronomy, University of Wisconsin Oshkosh, Oshkosh, 
       WI 54901, USA}  

\author{J. Borissova}   
\affil{Departamento de F\'isica y Astronom\'ia, Facultad 
       de Ciencias, Universidad de Valpara\'{\i}so, 
       Ave. Gran Breta\~na 1111, Playa Ancha, Casilla 5030,
       Valpara\'iso, Chile}

\author{J. R. De Medeiros}   
\affil{Departamento de F\'{i}sica, Universidade Federal do Rio Grande do Norte, 59072-970 
       Natal, RN, Brazil}     

\altaffiltext{1}{Current address: 
Las Campanas Observatory, Colina El Pino s/n, Casilla 601, La Serena, Chile}

\altaffiltext{2}{Current address: 
University of Heidelberg, Centre for Astronomy, M\"{o}nchhofstr. 12-14,
D-69120 Heidelberg, Germany}

\begin{abstract} %%% Abstract to run on from here.
We describe the status of a project whose main goal is to detect 
variability along the extreme horizontal branch of the globular cluster 
NGC~6752. 
%We have 
%produced deep time-series images of unprecedented quality for the cluster, 
%with which we were able to confirm the presence of variable candidates 
%in our studied fields, whose variability status are however still under 
%scrutiny. 
Based on Magellan 6.5m data, 
preliminary light curves are presented for some candidate variables. 
By combining our time-series data, we also produce a deep CMD of unprecedented 
quality for the cluster which reveals a remarkable lack of main sequence 
binaries, possibly pointing to a low primordial binary fraction. 
\end{abstract}

\section{Introduction}
Among field B-type subdwarf (sdB) stars, three types of non-radial pulsators 
have so far been detected, namely:

\begin{itemize}
\item EC 14026 (sdBV) stars: these are $p$-mode pulsators whose 
 temperatures fall in the range between 29,000 and 36,000~K. Their 
 periods are typically found in the range 100-200~sec, and their 
 amplitudes cover the range from 0.4 to 25\%. 

\item PG1716+426 (``Betsy'') stars: these are $g$-mode pulsators, with 
 temperatures in the range between 25,000 and 30,000~K. Their periods 
 are much longer, typically falling in the range between 2000 and 9000~sec, 
 and their amplitudes are smaller than 0.5\%. 
 
\item Hybrid stars: these are stars that present simultaneous $p$- 
  and $g$- mode oscillations. At present, only two examples have been
  reported in the literature \citep{abea05,roea05,ssea06}.
\end{itemize}

Performing asteroseismology on these stars holds the promise to unveil their 
innermost secrets, thus providing an exciting new route toward the solution of 
the so-called ``second-parameter problem'' \citep[][and references therein]{mc05}. 
The great promise of the technique notwithstanding, such variables have never 
been detected in previous searches in globular clusters \citep[e.g.,][]{mrea06}. 

Accordingly, the main purpose of the present study is to perform a new search 
for this type of variables in the relatively nearby southern globular cluster 
NGC~6752, which contains a very long blue HB ``tail,'' and thus many potential 
candidates for the three aforementioned variability types. 

\section{Observational Data}
Our study is based on data acquired with the Magellan II (Clay) 6.5m telescope
(259 $V$- plus 36 $B$-band images), taken in May 2006 at Las Campanas 
Observatory, Chile, using the MagIC high-resolution camera; and also on data 
acquired with the Magellan I (Baade) 6.5m telescope (829 $V$- plus 2 $B$-band 
images, taken in July 2007, using the IMACS mosaic camera in f/4 mode). 
While the MagIC images cover only two small fields of $2.4\times 2.4$~arcmin 
size, the IMACS images, taken in subraster mode, cover a much larger field, 
of size $15.4\times 7.7$~arcmin. Neither of these cameras is particularly 
suitable for rapid photometry. However, with a readout and exposure times 
both as short as 20~sec in the case of MagIC 
($\sim 50$~sec in the case of IMACS f/4 in subraster mode), there is   
a real chance to detect, for the first time, non-radial pulsators along 
the EHB of a globular cluster. Our ongoing search for variable stars using 
these datasets is based on both absolute photometry 
\citep[using {\sc Allframe/Trial};][]{pbs94} and difference imaging 
\citep[{\sc Isis} v2.2;][]{ca00}.

\begin{figure}[!ht]
\plotone{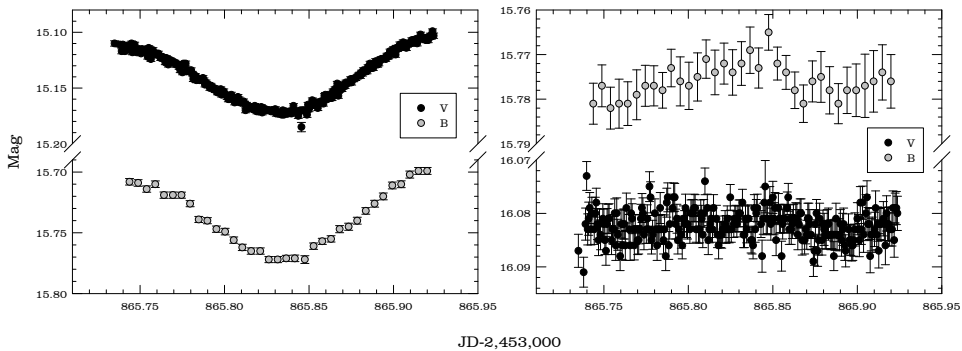}
\caption{Light curves for two variable star candidates in our studied MagIC
  fields. {\em Left:} an eclipsing binary. {\em Right:} a candidate variable
  on the blue HB.}
\label{vars}
\end{figure}

\begin{figure}[!ht]
\begin{center}
\includegraphics[scale=0.61]{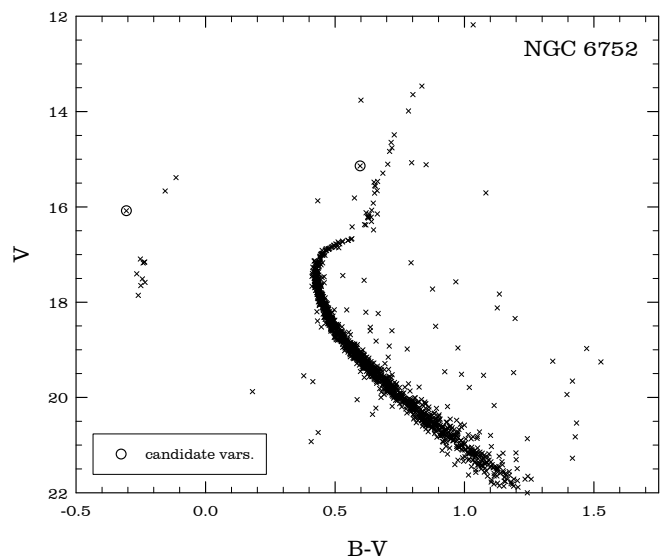}
\caption{Color-magnitude diagram for our studied MagIC fields. Encircled 
  symbols denote the candidate variables whose light curves are shown in 
  Figure~\ref{vars}.}
\end{center}
\label{CMDN6752}
\end{figure}

\section{Preliminary Results}

\subsection{Variable Star Candidates}
In Figure~\ref{vars} we show the first variable star candidates 
derived on the basis of our MagIC data. The star on the left is V9,  
which is likely to be a W~UMa-type eclipsing binary \citep{itea99} 
and does not fall on the HB (see Fig.~\ref{CMDN6752}). The star on the right, 
on the other hand, is a candidate variable that does fall 
on the blue extension of the HB. Analysis of the IMACS data should help 
clarify whether this is a real variable or whether the apparent variation 
seen in the data is due to systematic errors affecting our photometry. 

\subsection{Color-Magnitude Diagram}
Combining our time-series images using {\sc Allframe} provides us with a 
deep color-magnitude diagram (CMD) of unprecedented quality for the cluster 
(see Fig.~\ref{CMDN6752}, again based solely on our MagIC data). 
There is a remarkable lack of main-sequence (MS) 
binary stars outside the core of the cluster. This should be compared with 
a previous result by \citet{rb97}, who reported on a relatively large 
(15-38\%) MS binary population in the innermost regions of NGC~6752. Such 
a drop in the binary fraction with radius has recently been predicted in 
realistic $N$-body simulations by \citet{jhea07}, according to whom this 
implies a low {\em primordial} binary fraction as well. \citet{hrea07} have
recently obtained a similar result for the globular cluster NGC~6397. Such 
a finding may have important consequences for the formation of EHB stars 
in globular clusters vs. the field \citep{mc07}.

\section{Conclusions and Future Prospects}
Our analysis of time-series observations collected with the Magellan 6.5m 
MagIC camera already reveals a few intriguing variable candidates, including 
at least one on the hot extension of the HB, in two small fields located away 
from the cluster center. We have recently acquired extensive time-series 
observations using IMACS over a much larger field. These data will help 
confirm the nature of the suspected variables, and will presumably reveal 
a host of other faint/low-amplitude variables in NGC~6752. In addition, the 
new data will allow us to map the MS binary fraction in the cluster as a 
function of radius, which will reveal, for the first time, how in detail 
the binary fraction decreases as a function of radius in a globular star 
cluster. 

\acknowledgements Support for MC is provided by Fondecyt \#1071002. 
CMB was partially funded by Fundaci\'on Andes C-13798 
and the LOC. MA is supported by FONDAP 1501 0003. 
We warmly thank the LOC for the logistic arrangements that enabled 
us to present this paper at the conference.


\begin{thebibliography}{}

\bibitem[Alard(2000)]{ca00}
  Alard, C. 2000, \aaps, 144, 363

\bibitem[Baran et al.(2005)]{abea05}
  Baran, A., Pigulski, A., Koziel, D., Ogloza, W., Silvotti, R., \& Zola, S. 
  2005, \mnras, 360, 737

\bibitem[Catelan(2005)]{mc05}
  Catelan, M. 2005, preprint (astro-ph/0507464)

\bibitem[Catelan(2007)]{mc07}
  Catelan, M. 2007, preprint (astro-ph/0708.2445)

\bibitem[Hurley, Aarseth, \& Shara(2007)Hurley et al.]{jhea07}
  Hurley, J. R., Aarseth, S. J., \& Shara, M. M. 2007, \apj, 665, 707

\bibitem[Oreiro et al.(2005)]{roea05}
  Oreiro, R., P\'erez Hern\'andez, F., Ulla, A., Garrido, R., {\O}stensen, R., 
    \& MacDonald, J. 2005, \aap, 438, 257

\bibitem[Reed, Kilkenny, \& Terndrup(2006)Reed et al.]{mrea06}
  Reed, M. D., Kilkenny, D., \& Terndrup, D. M. 2006, Balt. Astr., 15, 65

\bibitem[Richer et al.(2007)]{hrea07}
  Richer, H., et al. 2007, \aj, in press (astro-ph/0708.4030)

\bibitem[Rubenstein \& Bailyn(1997)]{rb97}
  Rubenstein, E. P., \& Bailyn, C. D. 1997, \apj, 474, 701

\bibitem[Schuh et al.(2006)]{ssea06}
  Schuh, S., Huber, J., Dreizler, S., Heber, U., O'Toole, S. J., Green, E. M., 
    \& Fontaine, G. 2006, \aap, 445, L31

\bibitem[Stetson(1994)]{pbs94}
  Stetson, P. B. 1994, \pasp, 106, 250

\bibitem[Thompson et al.(1999)]{itea99}
  Thompson, I. B., Kaluzny, J., Pych, W., \& Krzeminski, W. 1999, \aj, 118, 
    462

\end{thebibliography}
\end{document}